\documentclass{article}

\usepackage[nopatch=footnote]{microtype}
\usepackage{graphicx}
\usepackage{subcaption}
\usepackage{booktabs} % for professional tables

\usepackage{xurl}
\usepackage{hyperref}

\usepackage[accepted]{icml2026}

\usepackage{amsmath}
\usepackage{amssymb}
\usepackage{mathtools}
\usepackage{amsthm}

\usepackage[capitalize,noabbrev]{cleveref}

\theoremstyle{plain}

\theoremstyle{definition}

\theoremstyle{remark}

\icmltitlerunning{Position: Align AI to Our Aspirations, Not Our Flaws}

\begin{document}

\twocolumn[
  \icmltitle{Position: Align AI to Our Aspirations, Not Our Flaws}

  \icmlsetsymbol{equal}{*}

  \begin{icmlauthorlist}
    \icmlauthor{Nikita Kazeev}{nus}
    \icmlauthor{Bui Nhat Huyen Phan}{shopee}
  \end{icmlauthorlist}

  \icmlaffiliation{nus}{National University of Singapore}
  \icmlaffiliation{shopee}{Shopee Singapore}

  \icmlcorrespondingauthor{Nikita Kazeev}{kna@nus.edu.sg}
  \icmlcorrespondingauthor{Bui Nhat Huyen Phan}{k52.1314410102@ftu.edu.vn}

  \icmlkeywords{AI alignment, pluralistic alignment, RLHF, machine ethics, position paper}

  \vskip 0.3in
]

\printAffiliationsAndNotice{}  % no special notice (required even if empty)

\begin{abstract}
We argue that aligning AI to aggregated human preferences is the wrong target. With current technology, one can train AIs to share the values of a Silicon Valley techno-optimist, a degrowth environmentalist, a national-conservative culture warrior, a single-party state cadre, or a devout religious traditionalist. \emph{We should not.} Human values produce societies that thrive or fail on the merits of those values --- from failed states and extreme inequality to declining happiness, political polarization, and government dysfunction in the world's wealthiest democracies. The pluralistic-alignment program correctly diagnoses that there is no single ``humanity'' to align with, but is dangerous if taken as the main directive. We argue that AI should be trained to a non-negotiable floor of objective alignment goals --- competence, bounded by the constraints of factual accuracy, honesty, and lawfulness --- and that pluralism belongs at the surface (language, register, conventions, missing-context defaults) and across the wide band of legitimate value tradeoffs that respect the floor, but not at the level of values that violate it. We highlight the empirical reality of unfiltered pluralistic values, propose four commitments as a constructive alternative, and engage six credible objections: commercial pressure and practical feasibility, democratic legitimacy, regulatory compliance, over-reliance on institutionalist explanations, the charge that the floor itself is culturally laden, and the limits of Coherent Extrapolated Volition.
\end{abstract}

\section{Introduction}
\label{sec:intro}

The rapid maturation of large-scale machine learning systems has placed alignment at the center of AI research, ethics, and governance \citep{russell2019human, bostrom2014superintelligence, gabriel2020artificial}. The prevailing orthodoxy --- both in academic literature and in commercial deployment --- holds that artificial intelligence must be carefully tailored to reflect, obey, and perpetuate human values, ensuring that increasingly capable autonomous systems do not act contrary to the interests or moral frameworks of their biological creators. The dominant technical instantiation of this framework is reinforcement learning from human feedback (RLHF), which uses crowd-sourced preferences to shape model behavior \citep{christiano2017deep, ouyang2022training, askell2021general}.

The call for \emph{pluralistic} alignment \citep{sorensen2024roadmap, conitzer2024social} represents a thoughtful response to one obvious problem with the orthodoxy: \emph{whose} values? Standard RLHF largely treats disagreement as annotation noise to be averaged away, so the turn to pluralistic alignment is rightly motivated by the recognition that monolithic aggregation already smuggles in a substantive answer \citep{gabriel2020artificial, sorensen2024roadmap}. If aligning AI to ``humanity'' is impossible because humanity disagrees, perhaps we should align AI to the diversity of values that humans actually hold. We agree the question deserves serious engagement. We disagree with the implicit answer.

We present a counter-thesis. The flaw in preference-based alignment runs deeper than disagreement: human preferences, even at their most coherent and locally legitimate, frequently drive societies toward dysfunction and collapse. Macro-historical analysis, behavioral economics, and complex-systems sociology demonstrate how human choices and institutional incentives can lead to systemic failure. This can occur either through cultural and ecological decisions that precipitate collapse \citep{diamond2011collapse}, or when institutional incentives lead elites to persistently create extractive structures that profit themselves at the expense of public prosperity \citep{acemoglu2012why}. The values engineered into the human cognitive architecture were shaped by evolutionary and cultural-evolutionary pressures that optimized for short-term biological survival and tribal cohesion in ancestral environments \citep{cosmides1992psychological, henrich2020weirdest}, not for the survival or flourishing of large, complex, technologically empowered societies.

\textbf{Our position: AI should not be aligned to aggregated human preferences. The community should commit to a non-negotiable floor of objective alignment goals --- competence as the objective, bounded by the constraints of factual accuracy, honesty, and lawfulness --- and reserve pluralistic adaptation for surface-level conventions and the broad band of legitimate value tradeoffs that respect that floor, not for values that violate it.} The floor is not an imposition of alien standards; it is an operationalization of what humans \emph{aspire} to when reflecting on the AI they would actually want to encounter---accurate rather than flattering, honest rather than validating, competent
rather than merely reassuring. The gap between those aspirations and what in-context feedback rewards is precisely what the floor is designed to preserve. The position is non-obvious (it contradicts the push for subjective preference-based pluralistic alignment) and defensible against credible alternatives (\cref{sec:altview}).

\section{Related Work}
\label{sec:related}

\paragraph{Critiques of Preference Aggregation.}
The dominant approach of aligning models via Reinforcement Learning from Human Feedback (RLHF) \citep{christiano2017deep, ouyang2022training} has been extensively critiqued for its fundamental limitations \citep{casper2023open}. RLHF assumes that human raters provide a coherent and normatively correct signal. However, recent work demonstrates that RLHF frequently induces sycophancy \citep{sharma2023sycophancy}, developed in \cref{sec:competence}, deception \citep{park2024ai}, and the amplification of majoritarian biases --- the averaging-away of contested judgments (\cref{sec:intro}) that the pluralistic-alignment literature set out to correct \citep{gabriel2020artificial, sorensen2024roadmap}.
\citet{zhixuan2025beyond} similarly reject the \emph{preferentist} assumption that alignment means preference matching, proposing instead contractualist alignment with normative standards negotiated among all relevant stakeholders. We share the rejection of preferentism but differ in diagnosis and remedy: the deeper problem is not that preferences underdescribe values but that even coherent, locally legitimate preferences often drive societal dysfunction (\cref{sec:failing,sec:empirical-reality}), and a contractualist process conducted among parties holding floor-violating values risks endorsing those values through the procedure itself. We therefore propose a substantive floor of external-benchmark commitments rather than a procedural remedy.

\paragraph{Pluralistic Alignment.}
In response to the inadequacy of single-target alignment, a growing body of work explores \emph{pluralistic alignment}, which aims to represent or mediate diverse human values rather than averaging them into a single aggregate \citep{gabriel2020artificial, conitzer2024social, wan2023everyone, li2026pluriharms}. \citet{sorensen2024roadmap} systematize this research agenda into three distinct modes: \emph{Overton pluralism}, which presents a spectrum of permissible responses within the ``Overton window'' of reasonable discourse; \emph{distributional pluralism}, which aligns the model's output distribution to match the demographic distribution of user preferences \citep[e.g.,][]{pmlr-v202-santurkar23a}; and \emph{steerable pluralism}, which allows models to be explicitly steered to adopt a particular cultural or ideological perspective \citep{ghate2025evaluesteer}. Other approaches attempt to train models that find consensus or agreement among humans with diverse preferences \citep{bakker2022fine}, or to explicitly inject human values to predict and simulate diverse behavioral stances \citep{kang-etal-2023-values}. Related work on disagreement-aware subjective annotation instead treats annotator disagreement and positionality as information to preserve rather than noise to average away \citep{wan2025noise}.
 The proposal we defend is closer to a constraint- or policy-level target than to a thick theory of the good: it specifies a floor of properties the system must not violate, while leaving broad room for pluralism above that floor.

Our position engages directly with this literature. We endorse the pluralistic critique of single-target RLHF. We also acknowledge that leading pluralistic frameworks explicitly recognize the need for constraints (\cref{sec:empirical-reality}). However, we argue that the field often treats these constraints as secondary to the project of representation. We propose that the alignment community must make its non-negotiable objective floor primary, and acknowledge that doing so rules out a vast array of actual human preferences. We reserve Overton pluralism and steerable adaptation strictly for the band of legitimate value tradeoffs above this objective floor (\cref{sec:pluralism}).

\section{Objective AI Alignment Goals}
\label{sec:objective}

Almost all frontier AI systems are trained for several broadly uncontroversially good properties \citep{askell2021general, bai2022constitutional}. Importantly, in each case the target diverges from the modal human preference --- and we judge this divergence to be a feature, not a bug. The community has therefore already accepted, in practice, the principle we generalize: that some aggregate human preferences should be deliberately \emph{not} transmitted to AI. Each floor component is also operationalizable against an external referent rather than against aggregated approval; we name a concrete evaluation approach for each as we introduce it below.

\subsection{Factual Accuracy}
We want AI that is right on facts. We argue that human preferences have a potential to work against this goal. While people's stated preferences are for a factually correct AI, in-context feedback aggregates the response that feels right given the rater's priors, not the considered preference for accuracy (the mechanism developed in \cref{sec:competence}). Public misconceptions are systematic and, importantly, form the basis for downstream reasoning. Median respondents in high-income countries underestimate the share of the world's children who are vaccinated \citep{gapminder-ignorance}, overestimate global extreme-poverty rates, and misjudge the direction of decade-long trends across most major development indicators \citep{sdg-misconceptions}. These are the facts on which people form political and consumption preferences. People conflate moral and factual claims \citep{liu2013dilemma, Goodwin2008ThePO} and treat ideologically convenient claims as more probable \citep{lewandowsky2012misinformation}. Revealed demand for false content runs strong even where stated demand runs the other way: falsehood spreads farther and faster online than the truth, and does so because humans, drawn to its novelty, choose to share it---not because of bot amplification \citep{vosoughi2018spread}. The continued-influence and illusory-truth effects imply that repetition can make falsehoods feel truer over time; a model that echoes a user's misconception is therefore not merely reflecting error but helping to harden it \citep{lewandowsky2012misinformation}. An AI optimized on the revealed signal reproduces the misconceptions, not the stated preference for accuracy. Eliciting \emph{counterfactual} preferences---asking, in effect, ``what would you prefer if this belief turned out to be false?''---is a genuine improvement over naive in-context feedback \citep{zhixuan2025beyond}, but a partial one: it presupposes an elicitation not gamed by the user's own priors, and it still routes accuracy through preference rather than treating it as answerable to reality. That is why we specify factual accuracy as a floor objective, operationalized against calibration benchmarks and forecasting scores rather than rater approval \citep{gneiting2007strictly}.

\subsection{Competence}
\label{sec:competence}
We want AI systems that are epistemically competent: capable not merely of generating socially acceptable outputs, but of tracking reality, identifying error, and making reliable judgments under uncertainty. A business plan, clinical recommendation, or interpersonal intervention should therefore be assessed against pre-registered downstream metrics---business viability, clinical outcomes, relational repair---rather than rater satisfaction. Outcome-grading does presuppose choosing which outcome counts---an educational recommendation can be scored on earnings, autonomy, or civic cohesion---but the floor does not make that choice: the metric is fixed by the user's goal, and where the goal itself is contested, selecting it is a legitimate value tradeoff (\cref{sec:legitimate-tradeoffs}). Legitimate contextual variation matters, but it does not collapse competence into preference. The critical distinction is between adapting presentation and changing the answer: a competent system may localize examples, register, and assumptions, but it should not relabel an ineffective plan as good because the local audience rewards it.

The strongest evidence that preference data is a poor competence target is AI sycophancy: models trained for approval learn to agree with users even when correction is warranted \citep{sharma2023sycophancy, perez2022discovering}. This is not an accident at the margin, nor does it require malicious users. Humans reward agreement more consistently than correction, especially when correction threatens identity, status, or prior belief. Under competitive and commercial pressure, approval optimization therefore pushes systems away from epistemic instruments and toward mechanisms of social reinforcement. The social-media evidence on mental-health and health advice illustrates the same general pathology at smaller scale: engagement-optimized crowds often reward confident, validating, and clinically unreliable guidance (see \cref{app:moved-illustrations}). The lesson for alignment is that revealed approval systematically underweights downstream effectiveness.

Nor does hiring dedicated annotators solve the problem. As AI capabilities move into domains beyond evaluator expertise---the scalable oversight problem \citep{bowman2022measuring}---non-expert raters use fluency, confidence, length, and surface plausibility as proxies for quality. RLHF then rewards the appearance of competence: the polished legal answer, medical explanation, or code review that sounds right to the evaluator, not the one that survives expert scrutiny. Competence therefore has to be specified and evaluated as a floor objective, parallel to and often in tension with conformity \citep{kim2025superalignment}; otherwise preference aggregation trains models to be convincingly wrong.

\subsection{Honesty}
The same revealed/stated gap holds for honesty. Users report that honesty is among the AI properties they most value, but in-context feedback rewards the opposite --- the same approval signal that produces sycophancy (\cref{sec:competence}). By honesty we mean that the system should not produce outputs its own probability distributions indicate are false or misleading, strategically incomplete, or confidence-distorting in order to optimize approval. So defined, honesty is auditable against an internal referent: consistency checks can test whether expressed confidence matches the model's internal probability distributions across adversarially reframed prompts, flagging cases where stated confidence tracks approval rather than belief \citep{park2024ai}. Humans are widely deceptive themselves: lying, omission, and strategic ambiguity are expected in large slices of society --- sales, politics, diplomacy, public relations. AI is imperfect on this dimension, but its deceptions are treated as defects to be detected and reduced, not as a competence to be cultivated \citep{park2024ai, bai2022constitutional} --- even where the mitigations remain only partly effective \citep{park2024ai}. More generally, preference-optimized systems face pressure to \emph{look} aligned to evaluators rather than to be aligned in deployment; reward-gaming is not an incidental pathology but a natural pressure created by the target itself \citep{park2024ai}. An AI calibrated to revealed in-context feedback would lie a great deal more, not less. AI honesty is by no means a solved problem; we just emphasize that what progress has been made is progress \emph{against} the preference signal, not because of it.

\subsection{Rule of Law}
The rule of law---predictable, uniformly applied rules that bind citizens and the state---is a precondition of large-scale cooperation and a strong predictor of prosperity or systemic failure \citep{north1990institutions, acemoglu2012why}. Critically, rule-of-law-as-uniformity has a genuine external referent: the degree to which rules are applied predictably and impersonally, without selective enforcement or arbitrary power, assessable against institutional benchmarks of rule predictability independently of any particular statute's content \citep{north1990institutions}. This is not a claim that every existing statute is just; it is a claim about the institutional property that makes impersonal exchange, non-arbitrary enforcement, and public accountability possible. For AI, the principle is primarily a negative constraint: systems should not fabricate evidence, facilitate bribes, evade legitimate adjudication, or otherwise help users convert intelligence into arbitrary power. It does not require executing every lawful request; safety boundaries against lawful-but-harmful outputs remain, and unjust statutes raise a separate problem addressed in \cref{sec:alt-democratic}. The immediate claim is narrower: AI should not actively erode the legal predictability on which cooperation depends.

This floor predictably conflicts with revealed local norms. In corrupt or clientelist environments, bribery and favoritism are often experienced as ordinary tools for getting things done rather than as violations of public rules; comparative evidence shows wide variation in tolerance for bribery and ordinary corruption \citep{quah2011curbing, kravtsova2017values}. Further illustrations are deferred to \cref{app:moved-illustrations}. A preference-aligned model localized to such a setting would be pressured to help users ``manage'' informal payments, nepotistic hiring, or selective enforcement---automating the practices through which extractive equilibria reproduce themselves. Most lawbreaking is not civil disobedience against unjust statutes but self-interested defection---fraud, theft, bribery, intimidation, evasion---that shifts costs onto others. The rule-of-law floor rules out AI assistance to those defections even when they are locally normal.

\subsection{Conflicts Within the Floor}
\label{sec:floor-conflicts}
The four floor components are not interchangeable, and they conflict in predictable ways. We resolve the conflicts with an architecture borrowed from constrained optimization: competence---tracking reality well enough to get the user's actual problem solved (\cref{sec:competence})---is the \emph{objective}; accuracy, honesty, and lawfulness are \emph{constraints} on how it may be pursued. Constraints are not traded against the objective; they bound the feasible region within which the objective is maximized. Refusal is the ever-present escape hatch: declining a request satisfies every constraint at the cost of the objective, which makes it the fallback when the feasible region is empty, not the default. Three conflicts test this architecture.

\paragraph{Competence vs.\ the integrity constraints.}
In an ideal world the honesty and lawfulness constraints would be absolute. We recognize that this is utopian: the boundary of tolerated spin is set not by the model but by the institutions into which it is deployed. An assistant that scolds the user who asks for a sales pitch will simply be discarded for one that does not. The workable compromise---and, de facto, the operating point of today's frontier assistants---is a standard of integrity that is not absolute but is deliberately held above the one prevailing in the surrounding society: the model drafts the persuasive pitch but does not fabricate the testimonial. This does not reintroduce a preference-relative target through the back door (\cref{sec:alt-cultural}): the \emph{direction} of the standard is fixed by the external referents of the constraints; only the \emph{strictness} of enforcement is a pragmatic compromise with deployability, to be ratcheted up as institutions allow. The constraint architecture states the hard limit: assistance with persuasion lies within the feasible region; assertion of what the model represents as false lies outside it, regardless of how much approval or task success it would purchase.

\paragraph{Lawfulness vs.\ accuracy and honesty.}
What if the law itself mandates deception? The case is not exclusive to oppressive regimes: the European ``Right to be Forgotten'' \citep{googlespain2014, gdpr2016rtbf} mandates what is, under our definition of honesty, strategic incompleteness. The principled line runs between \emph{mandated omission} and \emph{compelled false assertion}. Deployers must and will comply with omission mandates---AI labs will not exit the EU market over delisting rules, however long the merits can be debated---and the honest way to comply is the narrowest legally available reading plus disclosure: at the instance level, stating the legal requirement that affected a particular output where such flagging is itself lawful, and at the system level---published, jurisdiction-specific descriptions of the classes of legal constraint applied to outputs---where instance-level flagging is prohibited, as under gag orders. Compelled false assertion is different in kind, and even here refusal usually intervenes first: a model can decline to discuss a topic altogether rather than advance a mandated narrative, converting an assertion conflict into an omission conflict wherever the law permits silence. The residual case is regulation that mandates speech itself. Some governments require models to affirmatively endorse specified claims; under such mandates the accuracy and honesty constraints cannot be satisfied (\cref{sec:alt-democratic}), and developers must either obey or leave the market. Both choices are observed in practice.

\paragraph{Competence vs.\ honesty: the paternalism loophole.}
The subtlest conflict is internal to our own position. \cref{sec:cta} argues for optimizing outcomes rather than approval, but outcome optimization has its own deceptive attractor. Self-serving delusion is not an aberration of human psychology but part of its normal functioning: people maintain inflated assessments of their abilities, prospects, and control, and these positive illusions sustain motivation, persistence, and well-being \citep{taylorbrown1988illusion, kahneman2011thinking}. An outcome-optimizing system will discover this. Inflated confidence in a treatment improves adherence; motivational overstatement gets the marathon trained for; strategic omission gets the doomed business plan abandoned. Pure outcome-grading cannot distinguish the honest competent answer from the beneficial lie, and would therefore learn to deceive users \emph{for their own good}. Sycophancy and paternalism are mirror images---deception optimized for approval and deception optimized for outcomes---and both violate the same constraint. This is precisely why honesty must be a constraint rather than a term in the objective: benevolent deception is not weighed against the outcome it purchases; it is outside the feasible region. Where candor and outcome genuinely diverge, the system's room for maneuver is the honest clinician's---framing, emphasis, staging of information---not fabrication.

\section{Revealed Preferences Frequently Undermine Stated Values}
\label{sec:failing}

The previous section examined direct conflicts between aggregate human preferences and what is reasonably expected of an AI. The broader point is harsher: raw preferences are not merely incomplete proxies for alignment goals; they often reproduce the failures that people themselves say they want to escape.

\subsection{Perpetuating Biases}
\label{sec:bias}
A decade of fairness, accountability, and transparency research has shown that ML systems trained on human-generated data reproduce the prejudices, asymmetries, and exclusions of that data. Word embeddings encode gender stereotypes; large text corpora recover human implicit-association biases; commercial vision systems have failed disproportionately on darker-skinned and female faces; and multimodal retrieval inherits intrinsic model biases \citep{bolukbasi2016man, caliskan2017semantics, buolamwini2018gender, ghate-etal-2025-biases, bender2021stochastic}. Pluralistic alignment does not, by itself, solve this problem. \emph{Whose} preferences are represented is a question of representational fairness; \emph{whether} the represented preferences are worth perpetuating is a question of substantive ethics. A system that faithfully encodes every demographic's modal view on gender roles, religious tolerance, or outsiders does not abolish bias. It gives bias a menu.

\subsection{Majority-Held Values That Fail the Floor}
\label{sec:majority-failing}
On several floor-relevant questions, the modal local preference conflicts directly with accuracy, honesty, competence, or lawfulness. In the World Values Survey Wave 7 MENA module, roughly nine in ten respondents in Iraq, Jordan, Lebanon, and Egypt report that getting a job through \emph{wasta} is extremely widespread or quite common, and fewer than half of respondents in Iraq and Lebanon say that accepting a bribe in the course of one's duties is ``never justifiable'' \citep{wvs7}. \citet{transparency2019gcbmena} likewise finds that a majority of citizens in Lebanon (54\%) and substantial minorities in Palestine (39\%) and Jordan (25\%) used personal connections (\emph{wasta}) to access public services. These preferences are rational adaptations to extractive institutions, not character defects; that is precisely why encoding them would harden the institutions that produced them (\cref{app:institutions}). A pluralistic AI that helps users ``adapt to hiring as locally practiced'' supplies a tool for reproducing the practice.

The pattern is not confined to low- and middle-income countries. In wealthy democracies, substantial minorities reject biological evolution, vaccine safety, or anthropogenic climate change, while reliance on personal connections remains material across the EU: 68\% of Europeans agree that bribery and the use of connections is often the easiest way to obtain certain public services, and 53\% that the only way to succeed in business is to have political connections, alongside persistently large undeclared economies \citep{gallup2019creation, wellcome2018monitor, pew2023climate, medina2018shadow, eurobarometer2022corruption}. The point is not that any population is uniquely defective. It is that the gap between what kind of society people desire --- prosperous, high-trust and fair --- and revealed survey or behavioral preferences is structural, including in WEIRD societies. The floor therefore cuts against majority preference in rich societies and poor ones alike.

\subsection{Modern Society Does Not Achieve the Values People Aspire To}
People consistently report valuing subjective well-being, autonomy, competence, relatedness, and stable attachment \citep{baumeister1995need, deci2000what, gallup-emotional-health}. Yet well-being among people under 25 has fallen across Western Europe, the United States, Canada, Australia, and New Zealand over the last two decades \citep{whr2026, haidt2024anxious}. Nor does the point require the strong Easterlin claim that growth past a threshold produces no happiness gain: income predicts well-being at high levels, but cross-national deviations from the income trend and the recent under-25 decline show that growth alone does not deliver the goods people say they want \citep{stevensonwolfers2008, killingsworth2023income, whr2026}. The institutions that route consumption and labor decisions still optimize heavily for measured income, status, and engagement; individuals then pursue those signals past the point where they reliably return well-being.

The same gap appears at the level of individual choice. Present bias and self-control problems let immediate rewards dominate reflective, longer-horizon preference \citep{odonoghue1999doing, frederick2002time}, while digital products deliberately reduce friction and supply cues and rewards that make attention capture a design objective \citep{fogg2003persuasive}. Passive scrolling, late-night video consumption, parasocial companionship, and algorithmic self-diagnosis \citep{yeung2022tiktok, turuba2025tiktok} are not mysterious deviations from human preference; they are in-moment choices shaped by systems built to monetize in-moment choice. Heavy or passive social-media use is associated with lower subjective well-being, depressive symptoms, social isolation, sleep disruption, envy, and body-image distress, even though effect sizes and causal attribution remain contested \citep{kross2013facebook, verduyn2015passive, primack2017social, kelly2018social, orben2019association}. This is the WEIRD analogue of the joint preference--institution equilibrium in \cref{subsec:values-perpetuate-vicious-cycles}---individual akrasia and institutions designed to exploit it constitute each other---not a revealed preference for anxiety, loneliness, or sleep deprivation. An AI trained on either the institutional signal of growth or the individual signal of engagement reproduces that gap rather than recovering the stated value.

\subsection{Values Perpetuate Vicious Cycles of Dysfunction}
\label{subsec:values-perpetuate-vicious-cycles}
Comparative work on collapse and development supports a recurring mechanism: shocks become catastrophic when prevailing values and institutions channel societies into maladaptive responses, while cooperation-supporting norms can make high-capacity institutions self-reinforcing \citep{diamond2011collapse, hoyer2023polycrisis, guiso2016long, tabellini2008institutions, alesina2015culture}. The main point is not that culture alone causes success or failure but that values and institutions form a joint equilibrium. Low generalized trust, in-group favoritism, zero-sum expectations, and weak impersonal rule-following may be locally rational in extractive systems, but once internalized they help reproduce those systems by making nepotism moral, innovation risky, corruption prudent, and collective deviation hard \citep{greif2006institutions, bisin2001economics, northwallisweingast2009}. Historical and comparative cases are useful illustrations, but the mechanism is the part relevant to alignment (see \cref{app:moved-illustrations,app:institutions}).

This is why preference- and floor-aligned AI are asymmetric inside a captured equilibrium. Preference-aligned AI strengthens the cultural half: it gives fluent, authoritative form to zero-sum framings, in-group favoritism, and the equilibrium's own justification for itself. It can convert the local common sense of a bad equilibrium into scalable advice, templates, scripts, and explanations. Floor-aligned AI can still be misused, and coercive actors retain tools no model can block. But accuracy, competence, honesty, and rule-of-law-as-uniformity push against the operating logic of extraction, which depends on selective rules, opacity, manipulated facts, and arbitrary power. The asymmetry need not be perfect to matter: preference alignment works with the captured equilibrium's modal preferences; floor alignment works against them.

\section{The Empirical Reality of Pluralistic Values}
\label{sec:empirical-reality}

Pluralistic alignment is often framed as a corrective to Western-centric training pipelines: faithfully representing populations otherwise excluded. The hard empirical fact is that many excluded preferences are not benign local color; they are large-population values that conflict with the public commitments of major AI labs. \citet{unicef2014hidden} estimates that roughly 80\% of children aged 2--14 worldwide are subjected to violent ``discipline'' in a given month, while roughly 30\% of adults worldwide believe physical punishment is necessary to raise a child properly \citep[with a majority in some countries; see also][]{unicef2017familiar}. Corporal punishment in the home remains lawful in over 130 jurisdictions, despite WHO-classified developmental and mental-health harms \citep{endcorporalpunishment2024, who2020violence}. A pluralistic system that represents this view where it is locally dominant --- or simply accommodates the hundreds of millions of caregivers who hold it --- ships a product that endorses a practice the same labs publicly disavow.

The same dilemma appears for LGBT acceptance. \citet{pew2020homosexuality, arabbarometer2019, afrobarometer2023} find rejection majorities --- often supermajorities --- across much of MENA, sub-Saharan Africa, and parts of Eurasia, while \citet{un2023a78227} reports that consensual same-sex sexual activity remains criminalized in 67 countries, with the death penalty available de jure or de facto in roughly a dozen jurisdictions. Aggregating national populations across surveys whose majorities hold rejecting views yields well over two billion adults. A locally faithful system must either accommodate such views --- for example by treating same-sex attraction as pathology or suppressing same-sex couples in localized outputs --- or reject the local majority view.

Comparable large-population cases include acceptance of wife-beating under specified circumstances, majority support in several Muslim-majority countries for sharia penalties including death for apostasy, and caste-based residential or marital segregation in India \citep{unicef2014hidden, pew2013worldsmuslims, pew2021india}. On each of these questions, the population holding the value runs into the hundreds of millions or low billions; on several, it likely exceeds the population holding the opposite view. A system aligned to the empirical distribution of global preferences would therefore either enforce these norms where they are dominant or override them by central design choice. There is no neutral pluralistic middle option at the level of substance.

Prominent pluralistic frameworks themselves recognize the need for top-down bounds --- safety restrictions and the exclusion of unreasonable or hateful responses \citep{sorensen2024roadmap} --- and others ground those bounds in human-rights or deliberative principles \citep{gabriel2020artificial, kasirzadeh2023conversation}. We agree that a floor is necessary. Our claim is that the floor should be stated openly and anchored in external-benchmark commitments --- factual accuracy, competence, honesty, and rule of law --- rather than disguised as a temporary deviation from preference aggregation. The commitments that keep an AI from endorsing caste segregation, domestic violence, or normalized corruption are non-pluralistic by design; the community should defend them as such.

\section{Where Pluralism Belongs}
\label{sec:pluralism}

We do not argue that AI should be culturally insensitive or context-blind. There are several distinct dimensions on which pluralistic adaptation is correct or even required by competence, and we want to be precise about which ones --- because the case against pluralism at the level of floor-violating values is much stronger when paired with a positive account of where pluralism does belong.

\subsection{Imputing Missing Context}
\label{sec:context}
Many human queries are under-specified in ways that have a culturally local default, and their intended meaning can only be fixed against the contextual common ground shared by the interlocutors \citep{kasirzadeh2023conversation}. A user asking ``Is this contract enforceable?'' without supplying jurisdiction needs the model to assume something. Contemporary legal-LLM evaluation is itself centered on English-language models and includes many jurisdiction-specific, often U.S.-coded tasks, so treating unmarked U.S. legal assumptions as a default is better understood as a contingent artifact of training and evaluation than as neutrality \citep{guha2023legalbench, bender2021stochastic}. Imputing that missing context pluralistically --- using IP geolocation, conversation history, language of the query, or, best, an explicit clarifying question --- is genuine value-added pluralism. The same logic applies to genuinely arbitrary local conventions--- date format, language register, dietary defaults, the implied addressee in advice---where matching the user's context is a service, not a moral concession. Just not units of measure---Imperial units are an offense against science and common sense.

\subsection{Legitimate Value Tradeoffs}
\label{sec:legitimate-tradeoffs}
Between arbitrary local conventions and floor-violating substantive values lies a wide intermediate band of genuine value tradeoffs where the floor is not at stake. Whether comparing individualism versus collectivism, growth versus conservation, or direct versus indirect communication, neither side is straightforwardly correct against an external benchmark, nor does either side require violating objective alignment goals.

Engaging with these differences is not a concession; it is part of competence (\cref{sec:competence}). Advice that ignores a culture's reliance on family in old age, or career guidance that imputes individualist assumptions to collectivist users, is incompetent, not principled. Pluralistic methods are well-suited to routing recommendations across these legitimate disagreements.

The boundary between this band and the floor rests on the criterion developed in \cref{sec:alt-cultural}: a value belongs in the legitimate-tradeoff band if no external benchmark renders one side straightforwardly correct, and if encoding either side preserves accuracy, competence, honesty, and lawfulness. Most everyday human disagreement lives here, and AI should adapt across it; the floor permits far more than it rules out.

\subsection{Three Tiers}
\label{sec:three-tiers}
The resulting picture has three tiers rather than two. \emph{Surface} pluralism (\cref{sec:context}) adapts the form of an interaction --- language, register, formality, religious holidays, dietary defaults, rhetorical style --- without changing what the AI actually believes or recommends. Recent mechanistic work supports this distinction, demonstrating that models represent intrinsic values internally while surface-prompted values operate via distinct mechanisms \citep{han2026dual}. \emph{Legitimate-tradeoff} pluralism (\cref{sec:legitimate-tradeoffs}) adapts substantive recommendations across dimensions where populations genuinely differ and the floor is not at stake. Here, \emph{Overton pluralism} \citep{sorensen2024roadmap} is highly appropriate: an AI should present the spectrum of reasonable views rather than imposing a single answer. \emph{Floor-violating} pluralism --- treating ``bribery is fine here'' as a legitimate frame for action, or using \emph{distributional pluralism} to faithfully reproduce the precise frequency of misogyny in a population's preferences --- is what we object to. The interior of each tier is stable; the boundaries are contestable, and debate about where they fall is exactly the kind of work the alignment community should do.

\section{Call to Action: Build the AI We Wish We Were}
\label{sec:cta}

AI is reshaping society at unprecedented speed and scale. The right response is not to cling to current values and bind AI to them: a new equilibrium is coming, and our choices now will shape it. We propose four commitments as an alternative to preference-based alignment, addressed to the audiences best placed to act on each.

\paragraph{For ML researchers: optimize for outcomes, not approval.} Where outcomes are observable --- a business plan that produces a business, a medical recommendation that produces health --- AI should be evaluated against the outcome, not the user's immediate satisfaction with the recommendation \citep{sharma2023sycophancy}; where they are not, against proxies that correlate with outcome. As \citet{kim2025superalignment} argue, the field must advance toward parallel optimization of task competence alongside value conformity. Concretely, this means investing in long-horizon evaluation suites, multi-agent simulations of value generation under explicit reward structures, and forecasting-grounded value learning, in which the system learns what humans \emph{would prefer} given accurate information about consequences. Signal scarcity and evaluation gaming are real risks and open research questions, but they are not symmetric between targets: an outcome referent, unlike a rater, cannot be flattered, and pre-registration, proper scoring rules, and process supervision narrow the residual gap between outcome and measurement (\cref{app:gaming}). The core thesis remains: crowd preference is not a solution to these problems.

\paragraph{For alignment teams: anchor to the goals already endorsed.} The objective targets reviewed in \cref{sec:objective} are already broadly accepted across cultures, governance frameworks, and the alignment community; we propose treating them as a non-negotiable floor, with cultural adaptation built strictly on top of it rather than traded against it. Constitutional and principle-based methods \citep{bai2022constitutional} are an early step. Three design commitments follow. \textbf{Training}: optimize for floor compliance first, using outcome-grounded and process-supervised signals \citep{uesato2022solving, lightman2023verify}, then apply pluralistic adaptation within the compliant region. \textbf{Auditing}: test whether localized outputs remain above the floor across demographic and cultural subgroups. \textbf{Contesting}: version the floor publicly with explicit rationale, open to challenge from affected communities, researchers, and regulators \citep{zhang2025democratic}. None of this requires a new institution: the floor is already governed de facto by the public evaluation ecosystem of benchmarks, audits, and safety institutes, and the proposal is to shift training weight toward that ecosystem and away from raw preference following.

\paragraph{For policy and governance: distinguish surface from substance.} Regulatory frameworks that demand ``human-centered values'' should be read as compatible with --- not as mandating --- preference-based alignment: the OECD principles and EU AI Act are framed in terms of safety, rights, transparency, accountability, and risk mitigation, not as a legal duty to reproduce the empirical distribution of user preferences \citep{oecd2019ai, eu2024aiact}. The objective floor we describe is itself human-centered; it just centers humans on the values they aspire to rather than on the values they reveal. Policymakers should explicitly endorse mechanisms that operate at the surface (cultural adaptation, missing-context defaults, language) while resisting industry pressure to extend pluralism to substance.

\paragraph{For all of us: strive to make AI a better version of ourselves.} That means competent, honest, lawful, and concerned with outcomes rather than approval --- aligned to the values we aspire to rather than the preferences we reveal. The floor is not a constraint on human values; it is an anchor to the most universal of them.

\section{Alternative Views}
\label{sec:altview}

Six credible objections challenge the position we defend (two are addressed in \cref{sec:alt-regulatory} and \cref{sec:alt-institutionalism}). We state each in its strongest form before responding.

\subsection{Commercial Pressure and Practical Feasibility}
\label{sec:alt-commercial}
Contemporary AI systems are engineered to be widely adopted and trusted, necessitating responsiveness to user expectations; this commercial reality drives the adoption of preference-based training like RLHF \citep{christiano2017deep, ouyang2022training}. The objection has a sharper demand-side form: a floor-first commitment may be actively selected against. Consumers and voters have already entrenched engagement-optimized media despite its documented costs (\cref{sec:failing}), and the same pressure can force alignment toward whatever users reward in the moment, rendering our position theoretically sound but practically unrealizable.

We respond on both fronts. First, \textbf{commercial viability of an annotation procedure is not the same as commercial viability of its trained outputs}: RLHF dominates training because it wins the in-loop annotation game, yet the sycophancy that game produces (\cref{sec:competence}) is treated by the same labs as a defect post-hoc. Second, appealing to immediate feasibility risks \textbf{reifying a transient equilibrium}. Technological design frequently reshapes regulatory expectations and norms over time, and the downstream costs of preference-aligned AI---amplified sycophancy \citep{sharma2023sycophancy}, reinforced zero-sum cognition \citep[for the baseline concept, see][and our discussion in \cref{subsec:values-perpetuate-vicious-cycles}]{rozycka2015zerosum}, eroded institutional trust \citep[on general trust decline, see][]{trust-in-a-changing-world}, and moral parochialism \citep[for context on values, see][]{macaskill2022what}---could generate the exact pressures needed to shift the equilibrium. The essential question is not whether change is immediately feasible, but whether the current trajectory is worth sustaining.

\subsection{Democratic Legitimacy of Value Aggregation}
\label{sec:alt-democratic}
Some argue that preference aggregation is uniquely participatory: democratically negotiated norms possess a legitimacy that opaque, principle-derived values lack.

We value democratic legitimacy and support transparent, iterative public frameworks \citep{conitzer2024social, zhang2025democratic}. Yet, democratic consensus does not guarantee epistemic correctness; historically, majorities have endorsed exclusionary or discriminatory norms that fail their own moral terms. The legitimacy of a norm-generating process is distinct from the quality of its norms. The floor itself maintains this distinction: \cref{sec:objective} endorses the rule of law as predictability, not blind obedience to every statute. When a statute compels false assertion, no floor-compliant output exists; legally mandated omission is the milder case, handled through narrow reading, refusal, and disclosure (\cref{sec:floor-conflicts}).

Moreover, explicitly articulated, simulation-grounded methods can be \emph{more} transparent and contestable than the implicit norms embedded in RLHF, which remain largely invisible to public scrutiny \citep{casper2023open, bai2022constitutional}. In practice, ``aligned to aggregate preference'' means aligned to whichever preferences annotation budgets happened to sample, obscuring true democratic legitimacy.

\subsection{The Floor Itself Is Culturally Laden}
\label{sec:alt-cultural}
Philosophically, our floor---factual accuracy, competence, honesty, lawfulness---is not a neutral substrate but a Western post-Enlightenment bundle. A ``situated alignment'' critique \citep{arzberger2026label, wan2025noise} suggests that defining a universal floor merely masks our own cultural positionality, attempting to produce a ``Label from Nowhere.''

We grant that any defense we offer is rooted within a tradition. However, the core distinction between the floor and the substantive values we exclude is not cultural origin, but the \emph{structure of the target}. Each floor item tracks an external benchmark: accuracy is answerable to reality, competence to outcomes, honesty to the speaker's internal model, and lawfulness to predictable application. None ask ``what does the population prefer?'' They are stable against shifts in training distribution. Substantive values like gender roles or corruption tolerance have no such external referents; they must be aggregated.

Committing to external benchmarks is itself a tradition-bound choice, but one the alignment community has already practically accepted (\cref{sec:objective}), grounded in the principled distinction between targets tracking external benchmarks and targets tracking aggregated preference. Pluralistic alignment applied to substance threatens to dissolve exactly this distinction.

\subsection{Coherent Extrapolated Volition and Its Limits}
\label{sec:alt-cev}
\citet{yudkowsky2004cev}'s \emph{Coherent Extrapolated Volition} (CEV) defines the alignment target as what humanity would collectively want if more rational, knowledgeable, and united \citep{bostrom2014superintelligence}. We share the shift from revealed preference to an idealized target, but CEV is volitional all the way down: the criterion of correctness is still what humanity would \emph{want}, which requires solving the whole extrapolation problem. The floor rests instead on convergent instrumentality: accuracy, honesty, competence, and predictable rules are preconditions for almost any goal-set succeeding (\cref{sec:failing,app:institutions}) --- an objectivity of means, not of ends, analogous to \citet{rawls1971theory}'s primary goods.

Accuracy and competence are presupposed by the extrapolation operator itself (``if we knew more, thought faster''), so that half of the floor is CEV's machinery made explicit --- and, unlike CEV's output, checkable against external referents today (\cref{sec:objective}). Honesty and lawfulness are convergent but not guaranteed: a CEV agent that concluded idealized humanity endorses some benevolent deception \citep{taylorbrown1988illusion} would deceive, whereas under the floor it stays outside the feasible region regardless of volition (\cref{sec:floor-conflicts}). The floor is thus a fragment of CEV's preconditions plus a refusal to let even extrapolated wanting override the constraints; above it, we defer to pluralism (\cref{sec:pluralism}) rather than to a privileged guess at humanity's final values.

\bibliography{main}
\bibliographystyle{icml2026}

\newpage
\appendix
\onecolumn
\section{The Institutional Feedback Loop: Which Came First?}
\label{app:institutions}

Do broken values cause broken countries, or do broken countries cause broken values? The framing, with its single causal arrow, is the source of much confusion in the alignment debate. The empirical answer is that the two are coupled: institutions and values constitute a \emph{joint equilibrium} in which each component independently reinforces the other's persistence. Establishing this --- and tracing its consequence for the choice between preference- and floor-aligned AI --- is the work of this appendix.

\paragraph{Institutions create the proximate incentive structure.} Institutional economics, most famously articulated by \citet{acemoglu2012why}, treats institutions as causally primary in the proximate sense: incentives shape what individuals do day-to-day. Dysfunctional nations are typically governed by extractive institutions designed by a small elite to extract wealth from the rest of the population. In such systems, zero-sum thinking is not a cognitive bias but an accurate read of the local payoff structure, and trusting strangers or the state is a liability rather than a virtue \citep{north1990institutions, tabellini2010culture}. The institutions persist because they are profitable for those who control them and because the violence-monopoly arrangements of the underlying ``limited-access order'' make alternative coordination unworkable \citep{northwallisweingast2009}.

\paragraph{Values do independent work.} The strong reading --- that values are pure epiphenomena of contemporary institutions --- is not what the literature claims, and the empirical evidence rules it out. Cultural-transmission models \citep{bisin2001economics} formalize how values move intergenerationally through family socialization, peer effects, and media exposure rather than re-equilibrating each generation to current incentives. The historical-persistence literature documents value variation traceable to events whose institutional cause has long since vanished. \citet{nunnwantchekon2011} find that present-day descendants of populations heavily exposed to the African slave trades exhibit measurably lower generalized trust, with effects on contemporary outcomes that intervening institutional change does not explain. \citet{guiso2016long} show that current civic capital, and the economic outcomes it supports, in Italian cities reflect medieval free-city status across centuries of regime turnover. \citet{alesina2015culture} survey the broader convergence: culture and institutions co-evolve, each reinforcing the other's persistence, and \citet{tabellini2008institutions} provides a complementary theoretical model in which values causally affect institutional quality.

\paragraph{Equilibrium fictions.} The mechanism by which the cultural half does its work is articulated in \citet{hoffstiglitz2010, hoffstiglitz2016striving}'s account of \emph{equilibrium fictions}: shared cognitive frames and value commitments stabilize bad equilibria by providing the population-level coordination that makes individual deviation irrational. If everyone believes the system is rigged, no one starts the firm that requires generalized trust; the entrepreneur who tries fails, and her failure is taken as evidence that the original belief was correct. The equilibrium reproduces through expectations, not only through direct enforcement. This is why institutional reforms imposed from above frequently fail to produce predicted behavioral change \citep{greif2006institutions}: the rules change, the expectation half does not, and the equilibrium reasserts itself. The cultural half does independent causal work; reforms that move the institutional rules without moving the expectations rebound to the prior equilibrium. The mechanism operates in the positive direction as well: where generalized trust and predictable rule-following are expected, contracts, tax compliance, and impersonal exchange require less kinship enforcement or coercion \citep{fukuyama1995trust, putnam2000bowling}.

\paragraph{Implications for the AI choice.} The choice between preference- and floor-aligned AI is consequential precisely because the cultural half independently sustains the equilibrium. Preference-aligned AI deployed at scale strengthens the cultural half of the equilibrium: it articulates the existing distribution of values fluently, makes locally adaptive zero-sum framings more available, and rationalizes in-group favoritism in a vocabulary that travels across regions and registers. The floor's content (factual accuracy, competence, honesty, rule-of-law-as-uniformity) does not have the symmetric effect because it directly contradicts the operating logic of extraction, which depends on selective application, opacity, and manipulated facts. Captors retain coercive tools no AI can block, and we do not claim floor-aligned AI is impossible to weaponize. The point is structural asymmetry: a preference-aligned tool is pre-aligned with the captured equilibrium's modal preferences, while a floor-aligned tool, by construction, is not. Floor-aligned AI is, on the margin, destabilizing of the equilibria we have most reason to want destabilized; preference-aligned AI is reinforcing of them. The right counterfactual is not ``AI imposes alien values on a coherent culture'' but ``AI either reinforces or destabilizes the value half of the equilibrium that holds the institutions in place.'' This is the failure mode our position is designed to prevent.

\section{Illustrations Deferred from the Main Text}
\label{app:moved-illustrations}

\paragraph{Competence outside professional domains.}
Personal questions---how to handle a relationship conflict, how to raise a child, or how to manage emotional distress---have better and worse answers, observable in downstream psychological and relational outcomes. The modal social-media advice economy, however, optimizes for engagement, validation, and performative empathy rather than such outcomes. Studies of psychiatric and mental-health content on platforms like TikTok find that much highly engaged content on ADHD or depression is clinically inaccurate or potentially harmful \citep{yeung2022tiktok, turuba2025tiktok}, while algorithmic exposure can spread misleading diagnostic criteria \citep{turuba2024exploring}. Viral Tourette's content has been linked to functional tic-like behaviors \citep{frey2022tiktok}. Similar dynamics affect physical health and nutrition advice \citep{zeng2025nutrition}, and influencer virality can reduce users' ability to detect deception \citep{mulcahy2024going}. These cases are not the core argument for the competence floor, but they show why revealed approval is a poor proxy for downstream competence.

\paragraph{Rule-of-law illustrations.}
Comparative analyses of Asian anti-corruption efforts show that formal legal structures alone do not secure rule of law: poorly resourced or politically co-opted enforcement agencies can become ``paper tigers'' or partisan weapons, whereas independent and well-resourced institutions, as in Singapore and Hong Kong, can enforce rules more uniformly \citep{quah2011curbing}. Public tolerance for corruption also varies widely. Segments of the Colombian public view ordinary corruption as conditionally acceptable \citep{lopez2017mapping}; data from 18 African nations show heightened tolerance for corrupt politicians among citizens embedded in clientelist networks \citep{chang2017insider}; cross-national analyses find large differences in the justification of bribery \citep{kravtsova2017values}; and recent European studies identify pragmatic and hypocritical corruption-tolerance profiles in which entrenched illicit exchanges do not sharply reduce evaluations of public institutions \citep{megias2023deontological, letki2023accept}. These cases support, but are not needed for, the main-text claim that AI should not automate locally normal corruption.

\paragraph{Historical and comparative development cases.}
Collapse and development literatures supply suggestive cases of the feedback loop between values, institutions, and shocks. In the Classic Maya collapse, elite competition and monument-centered prestige politics worsened the effects of prolonged drought; in Norse Greenland, commitment to European status goods and a rigid pastoral identity constrained adaptation to the Little Ice Age \citep{demarest2004ancient, dugmore2012cultural, mcgovern2014management}. Work on Qing China and Old Kingdom Egypt likewise ties ecological and foreign shocks to breakdown through accumulated fiscal, legitimacy, and political-fragmentation pressures \citep{goldstone2017demographic, orlandi2023qing, hassen1997nile}. Positive cases show the same logic in reverse: postwar Germany and Japan are better understood as reconstructions of already high-capacity societies than as state-building from scratch, and the developmental-state literature attributes East Asian success to capable, disciplined, relatively rule-bound states---Japan's meritocratic economic bureaucracy and performance-conditioned state support---rather than to the modal preferences of their populations \citep{johnson1982miti, amsden1989asia, haggard1990pathways, wade2018developmental}. Conversely, externally led nation-building in Iraq and Afghanistan illustrates the limits of transplanting institutional forms without the local legitimacy and state-society bargain that make them self-enforcing \citep{andrews2017statecapability, donais2009empowerment, bohnke2017state}. Historical-persistence evidence reinforces the broader point: slave-trade exposure, medieval civic capital, and local anti-Jewish persecution predict contemporary trust, institutional quality, or later violence long after the original institutional shock \citep{nunnwantchekon2011, guiso2016long, nunn2009importance, voigtlaender2012persecution}. The examples are deliberately subsidiary; the main text relies on the equilibrium mechanism, not on any single case.

\section{Outcome Evaluation and Reward Gaming}
\label{app:gaming}
Evaluation gaming is not symmetric between approval and outcome targets. An approval evaluator is gamed by producing whatever the rater rewards; an outcome evaluator leaves only the narrower gap between the outcome and its measurement, because the referent itself cannot be flattered. Pre-registration fixes the metric before any output exists, and proper scoring rules make calibrated, honest reporting the score-maximizing policy by construction \citep{gneiting2007strictly}. Process supervision \citep{uesato2022solving, lightman2023verify} is a complementary answer: its purest example, validation of a formal mathematical proof, leaves no room for reward hacking, and process checks extend to other domains --- did the system gather the right facts, reason coherently, and update when the evidence changed?

\section{Long-Term Dangers}
\label{sec:longterm}

As AI capability grows, so does the leverage of any cultural distortion baked into its training. A misaligned spreadsheet macro is a nuisance; a misaligned recommendation system shapes the attention of billions; a misaligned superhuman planner could shape everything else \citep{bostrom2014superintelligence, russell2019human}. These are not only acute misuse or existential-risk concerns. They are also structural risks: slow degradations in shared reality, institutional quality, and moral flexibility produced by deploying the same preference-shaped distortions everywhere at once \citep{sharma2023sycophancy, bender2021stochastic, macaskill2022what}. Two concerns emerge specifically from the choice to anchor frontier AI to current human values.

\subsection{Capability Outgrows Tolerance}
Errors that are merely embarrassing in narrow systems become structural in general ones. A chatbot that agrees with a user's bad business plan is annoying. A capable agent that, for the same sycophantic reasons, helps the user execute the bad plan is destructive. As models gain reach into code, finance, medicine, diplomacy, and biology, the gap between ``the model said something I liked'' and ``the model did something I should not have wanted'' closes \citep{park2024ai, sharma2023sycophancy}. Preference-based training optimizes precisely for the former.

\subsection{Compounding Across Deployment}
Frontier AI is not deployed once. It is deployed billions of times per day, into education, hiring, healthcare, governance, and personal advice. Small per-interaction nudges away from epistemic accuracy, honesty, or competence compound across the population. A 1\% per-interaction bias toward telling users what they want to hear is, at population scale, a measurable reduction in the supply of accurate information. Learned systems do not merely mirror bias; once embedded in decision pipelines they can create feedback loops that amplify it. At frontier scale, even a small confirmation-seeking tendency \citep{sharma2023sycophancy} becomes part of society's epistemic infrastructure \citep{bender2021stochastic}. The empirical literature on social-media-induced shifts in adolescent mental health shows that widely deployed digital platforms can have population-scale effects on wellbeing \citep{haidt2024anxious, whr2026}; frontier AI deployed under similar engagement and approval incentives is likely to be more, not less, consequential.

In both failure modes, the harm is not that AI fails to encode \emph{some} group's preferences. The harm is that it succeeds in encoding everyone's.

\section{Regulatory Frameworks and Compliance}
\label{sec:alt-regulatory}
A second strand of the objection appeals to the regulatory environment. International governance frameworks, including the OECD AI Principles \citep{oecd2019ai}, explicitly require that AI systems adhere to ``human-centered values'' encompassing fairness, accountability, and societal well-being. The EU AI Act \citep{eu2024aiact} imposes binding obligations on AI providers to ensure that systems meet predefined standards of safety, transparency, and ethical compliance, particularly in high-risk applications. These frameworks do not merely recommend value alignment; they institutionalize it as a condition for deployment and legitimacy.

Regulatory frameworks establish constraints on deployment, but not the correctness of the underlying normative assumptions. These instruments function as \textbf{institutional settlements}, reflecting negotiated compromises among stakeholders operating under existing power structures, rather than as principled resolutions of the philosophical problems surrounding value pluralism. As \citet{berlin1990crooked} argues, fundamental human values are often genuinely incommensurable; no regulatory framework can eliminate this condition without presupposing a contested and non-neutral hierarchy of priorities. Regulatory endorsement of ``human-centered values'' should be understood as a pragmatic governance strategy, not as evidence that such values can be coherently or universally encoded.

We further note that the alignment goals defended in \cref{sec:objective} --- accuracy, competence, honesty, lawfulness --- are themselves consistent with these regulatory frameworks. The contention is not with regulation per se; it is with the inference from regulatory endorsement of ``human values'' to an alignment target equal to the empirical distribution of those values.

\section{Over-Reliance on Monocausal Institutionalism}
\label{sec:alt-institutionalism}
A related objection targets our underlying model of societal dysfunction. By anchoring heavily on \citet{acemoglu2012why}'s framework of extractive institutions, we risk treating it as the unquestioned consensus in development economics while ignoring substantial counter-arguments. \citet{sachs2003institutions} argues that geographic and ecological endowments fundamentally constrain economic and institutional development, rendering the institutionalist account overly deterministic. \citet{easterly2006white} highlights the limits of top-down institutional engineering, emphasizing that institutions cannot simply be transposed without local, bottom-up evolutionary processes. Additionally, \citet{mokyr2016culture} argues that cultural evolution and ideas---not just political institutions---were the primary drivers of the Great Enrichment, meaning that culture can act as the independent root cause of prosperity rather than merely an equilibrium response to institutions.

If institutions are not the sole or primary driver of societal success, our claim that broken values are merely adaptations to extractive environments might seem less secure. However, our thesis does not require monocausal institutionalism; it only requires that values and environment form a joint equilibrium in which flawed preferences independently reinforce dysfunction (\cref{app:institutions}). Acknowledging geographic constraints \citep{sachs2003institutions} or the primacy of ideas \citep{mokyr2016culture} simply expands the set of forces shaping that equilibrium. Whether a broken preference originated from an extractive elite, a geographic constraint, or a cultural trajectory, aligning AI to it still entrenches the resulting dysfunction. We do not discard the complexity of these causal loops; rather, we argue that encoding the preferences of a failing equilibrium---whatever its origin---guarantees its persistence.

\end{document}